# Meaningful Human Command: Towards a New Model for Military Human-Robot Interaction

Adam J. Hepworth[1], Zena Assaad[2], Austin Wyatt[3] and Hussein A. Abbass[4]

## Abstract

Military human robot interaction (MHRI) presents a novel opportunity to blend the capabilities of autonomous and Artificial Intelligence (AI)-enabled systems with the skills and expertise of humans. The concept promises military advantages and greater operational effectiveness and efficiencies. However, the associated human-AI dynamics create challenges when attempting to design, implement, and operationalise the increasingly symbiotic relationship between humans and machines. Meaningful human control (MHC) is a popularised conceptualisation of what is deemed a responsible interaction among human and artificial agents; however, this notion falls short in military contexts and hinders the realisation of military advantages that could be achieved by advancing the adoption of responsible AI. This paper presents meaningful human command (MHC1) as a more operationally effective concept for advanced military command and control systems that embed AI-enabled autonomous systems. We introduce, explore, and unpack meaningful human command in the context of military human-robot interaction, presenting a vignette that offers a technologically feasible concept of an AI-enabled system within military operations. The vignette is used to guide, contextualise, and add realism to the narrative describing the concept and highlights associated MHRI challenges.

## Key points/Objectives

- Introduce the concept of Meaningful Human Command (MHC1).
- Describe Human-Robot Interaction (HRI) from a military operational perspective.
- Understand key considerations for technologists (HRI professionals) and practitioners (military and policy) stakeholders working on military HRI challenges.
- Propose MHC1 as a framework to address implementation concerns with MHC, while aligning with military mission command implementations.

---

[1] University of New South Wales.
[2] Australian National University.
[3] RAND Australia.
[4] University of New South Wales.

## Introduction

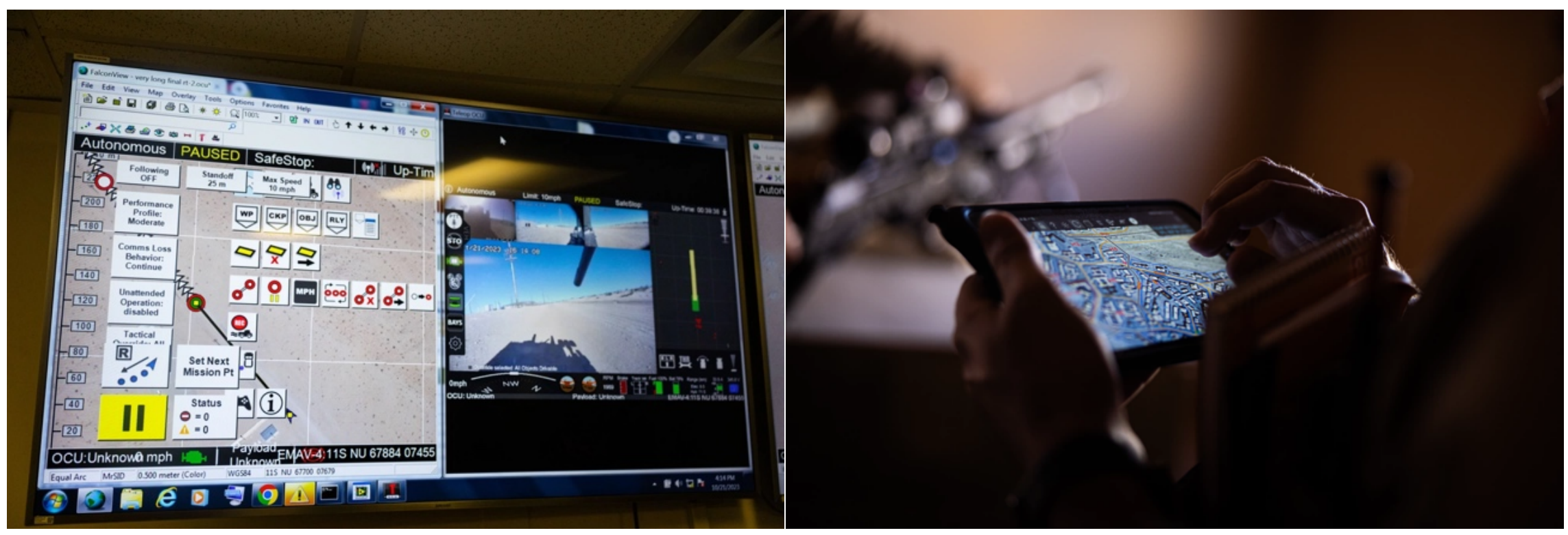

*Figure 1: Contemporary command and control interfaces for military human-robot interaction (Left – LCpl Joshua Simmonds/US Department of Defense; Right – Sgt Makayla Elizalde/US Department of Defense).*

Military Human-Robot Interaction (MHRI) sits at the intersection of the military as a profession of practice and Human-Robot Interaction (HRI) as a field of scientific inquiry (Möller et al., 2021). Work at this intersection falls into two broad categories. In the first category, knowledge from non-military domains can be transferred to the military domain without substantial modification to the HRI knowledge base. For example, searching for weeds in agricultural farming situations or survivors in a humanitarian disaster setting that share similarities with surveillance operations in military settings (Pavithra and Kavitha, 2022), as is seen throughout the conflict in Ukraine today (Santora et al., 2025).

The second type of category knowledge is where a knowledge transfer is insufficient to address the uniqueness of the military domain, requiring the creation of new knowledge to meet operational requirements. Transitioning a surgical robot from civilian to military hospitals, under harsh operating conditions, will introduce significant challenges for the robot and stressors for the humans within the HRI dynamic. These challenges sit well-above those used to design, develop, operationalise, and test medical robots operating in benign civilian contexts. Such challenges to overcome include constrained communication bandwidth, latency, and intermittent signal stability, among others (Rosen et al., 2011).

Humans and robots in situations that fall within this second category operate in harsh environments where decisions are made in the presence of unique stressors, and the preferences and priorities of decision makers could yield significantly different outcomes from the same input information. Many novel human factors issues across research areas of interaction, cognitive workload and understanding are emerging within the military domain due, in part, to these challenges (Sheridan, 2016a). The physical (terrain, weather, injuries, robot and human physical malfunctions), psychological (psychological stressors, sensory impairment factors, involuntarily psycho-physiological changes, and rates of depletion in cognitive resources), and social (human-to-human and human-to-non-human relationships) factors that change behaviours in war situations, have flow-on effects to broader military operating structures, including the command structure. A team leader could die, a subordinate could be deliberately or accidentally isolated from the team's purpose, and robotic capabilities could increasingly be assigned to a soldier to control or even command, particularly in settings deemed too dangerous for humans (Sheridan, 2016a).

An increased or imbalanced robot-to-human ratio, an increase in the levels of autonomy of robotic systems and the delegation of complex actions to robots afford these capabilities substantially greater degrees of freedom in the presence of extreme, severe and unpredictable operational stressors. Many scholars have discussed the need for meaningful human control (MHC) to provide principles and guidance for MHRI requirements across design and testing (technical), employment (procedural use) and legal (policy) aspects (Horowitz and Scharre, 2015). Philosophers argue that MHC is a topic of scientific inquiry, while technologists and engineers approach it through the practicalities of shared situational awareness, human-robot interfaces, communication networks, autonomy levels, cyber-security, and system-of-systems assurances. The ongoing discourse on the concept of MHC, which has yet to reach consensus, has led to changes in how MHC is defined and considered. More recently, the Group of Governmental Experts on Lethal Autonomous Weapon Systems (GGE LAWS) has transitioned away from MHC and now adopts the term “context-appropriate human judgement and control” (UNODA, 2025). Aligned with this evolving discourse, we posit in this chapter that MHC is a limiting concept; as robot autonomy increases and the role of the human within the robotic operational context moves from control to command, MHC constrains the operating envelope of HRI. It assumes low-level machine autonomy and a micro-management approach to HRI.

Central to the challenges in this second category of MHRI is the command and control (C2) system, the connective tissue that brings humans and robotic systems together in a variety of teaming arrangements and with the supporting infrastructure to achieve mission objectives. In this chapter, we highlight the limits of MHC and propose meaningful human command (MHC1)[5] as a viable evolution of the concept.

In the remainder of this chapter, we review selected pieces of the literature that we have identified as relevant and foundational for this chapter. We then present the characteristics of MHRI and a few key definitions. A vignette is then presented, serving two purposes. One is an illustrative contextualisation of the concepts and challenges covered in the chapter. Second, as a context for HRI and MHRI scholars to use in future research. We conclude the chapter with a discussion to summarise lessons learnt and key considerations for both HRI researchers and military practitioners.

This chapter encapsulates terminology and concepts that are subject to varying interpretations across disciplines. To alleviate ambiguity, we first provide definitions of key terminology in the following section.

## Definitions

One ongoing challenge in the military applications of artificial intelligence (AI) and robotics is the continued uncertainty around essential terminology and standards. Such clarity is necessary for advancing from conceptual agreement to meaningful legal, doctrinal, and normative structures. In the absence of universally agreed definitions, it is useful to devote a section of this paper to outlining the authors' perspective and approach to key concepts.

Beginning with the foundational technology, definitions of AI are prominent but varied in the literature.[6] The definition of AI that we adopt within this paper is 'Artificial Intelligence is the automation of cognition' (Abbass, 2021). This concise definition offers simplicity that allows AI to be generalised beyond human intelligence. It centres AI on knowledge, the process to identify, acquire, process, and generate knowledge, and knowing how to know.

Diversity in the style and purpose of definitions for AI exists, for instance, the Organisation for Economic Co-operation and Development (OECD) defines it as "a machine-based system that, for explicit or implicit objectives, infers, from the input it receives, how to generate outputs such as predictions, content, recommendations or decisions that can influence physical or virtual environments. Different AI systems vary

---

[5] Our shorthand for C1 refers to the first C in 'Command and Control'. We do not suggest meaningful human command and control (MHC2) due to the confusion such a terminology could create among both military and HRI professionals.

[6] Commonly cited definitions include the US Department of Defense AI Strategy, the definition contained in 15 U.S.C. 9401, which was also used in White House Presidential Action Memorandum of January 23, 2025, *Removing Barriers to American Leadership in Artificial Intelligence*, the definition adopted by the National Defense Authorization Act (NDAA) FY2019, and the definition adopted by the Organization for Economic Co-operation and Development (OECD).

in their levels of autonomy and adaptiveness after deployment" (European Commission, 2018).

Another definition by the United States (US) Department of Defence (DoD) 5000.101 considers AI as "a machine-based system that can, for a given set of human-defined objectives, make predictions, recommendations, or decisions influencing real or virtual environments. AI systems use machine- and human-based inputs to perceive real and virtual environments; abstract such perceptions into models through analysis in an automated manner; and use model inference to formulate options for information or action." This definition focuses on how AI works and the inner process of AI systems (Department of Defense, 2024).

Published in the Australian Army Robotic and Autonomous Systems Strategy v2.0, AI is defined as "a collection of techniques and technologies that demonstrate behaviour and automate functions that are typically associated with, or exceed the capacity of, human intelligence" (Australian Army, 2022). This definition reflects that AI is a foundational technology, more analogous to electricity or the internet than to a specific artefact, and that it draws on a variety of underpinning techniques and technologies. It also importantly notes that AI systems can be purely software systems, or cyber-physical systems when embedded into physical artefacts or hardware devices. (Australian Army, 2022).

AI systems do not always hold decision-making functions in physical or virtual environments. For example, an AI system that analyses a stream of data from diverse sensors to present commanders with a common joint situation awareness picture does not directly recommend or make decisions about courses of action. One would see this AI system still influencing human cognition through information sharing, which could, in turn, influence the decisions made. However, this could be said about every piece of information assessed by a human and its potential influence on human decision-making. Therefore, we argue that some AI systems may be purely descriptive, without predictive output or recommendations and may not need to influence physical or virtual environments to be considered AI.

AI is a categorical umbrella term that encompasses a range of applications of statistical reasoning-based models (Bishop, 2021), which include, for example, large language models, computer vision models, and deep neural networks. While each of these models falls under the broad concept of AI, they are not the same. Much of the discourse around AI centres on the broader categorical label of AI, omitting vast detail and application specificity.

Among that broad discourse is the emerging concept of Responsible AI (RAI), which reflects a deliberate integration of ethical and legal principles into the development and use of AI-enabled systems. The goal here is to achieve a workable balance between military effectiveness, technological limitations, and minimising adverse ethical risks. RAI policies have been adopted by a range of states, including the European Union, the Republic of Korea, Australia, the United Kingdom and the US governments, amongst others. Whilst the specific definitions vary, such policies generally emphasise that an RAI-enabled system should be ethical, equitable, traceable, explainable, and governable (for instance, see: (Steckman et al., 2025), (Blanchard et al., 2025), and (Jafari Meimandi et al., 2025)).

These characterisations have come to represent growing concerns about the opacity of AI-enabled systems, particularly in safety-critical contexts such as the military. Where RAI approaches have been accompanied by an emerging recognition of the alternative military uses of AI-enabled technologies (for example, logistics, decision support systems, and training), autonomous weapon systems remain both central and most controversial within the scholarly, governmental and intergovernmental discourse. Yet, this application remains without a universally accepted definition, standard, or benchmark for objectively determining whether a given artefact can be classed as a fully autonomous weapon system.[7] The most commonly cited definition is advanced in the 2023 update of Department of Defense Directive 3000.09, which focuses on whether a system "can select and engage targets without further intervention by an operator" (Department of Defense, 2023). Similar definitions have been adopted globally, although there are differences in detail and scope. Additionally, it is generally accepted in the literature that autonomy is not a binary characteristic; rather, it represents a spectrum of capabilities of a system's capacity to self-direct its movement and other critical functions (such as sense, identify and engage targets) (NATO, 2023).

## Literature Review

The intersection of military C2 and HRI warrants its own field of inquiry. The implementation of robotic systems in military operations is paralleled with the operational challenges of this engineered scale. This section is arranged into three sub-sections. In the first, we present a high-level summary of military C2, focusing our attention on mission command. We explore HRI research in the second sub-section, then briefly discuss adjacent fields such as Human-AI Teaming and Human-Machine Teaming, concluding with a review of typical research applications. In the final sub-section, we use these two lenses to characterise HRI in the military domain, explore the uniqueness of this underserved research area and frame the perspective of this work.

[7] The value of definitional clarity and objective standards was highlighted in 2018, when China publicly supported a legal ban on Lethal Autonomous Weapon Systems, but based that support on a set of technologically and logically implausible criteria.

## A brief introduction to military command and control

Militaries have practised various forms of Command and Control (C2) for thousands of years, with approaches ranging from highly centralised to more distributed models for planning, coordinating and executing military activities (Alberts and Hayes, 2006). Research into methods of C2 has increased substantially through the 21st century, commensurate with an increasing complexity of military operations and capabilities (Alberts and Hayes, 2006).

The function of C2 is often characterised as an information-push process (Builder et al., 1999), in which structured and specific information is conveyed to and from personnel for a well-defined purpose. Focusing on its underlying concept, C2 can be broadly understood as the exercise of command to achieve military objectives (Builder et al., 1999).

Doctrinally, a distinction between 'command' and 'control' exists, with the application of command and control operating on a spectrum between these two extremes in practice. For example, the Australian Defence Force Philosophical Doctrine (Australian Defence Force, 2024) on Command states that 'there are five elements that constitute the nature of command. These are responsibility, authority, decision-making, leadership and accountability. Three of these elements are what the commander has: responsibility, authority and accountability. The remaining two are the essence of what a commander does: makes decisions and leads.'

This philosophical doctrine goes on to state that control is 'the authority exercised by a commander over part of the activities of subordinate organisations, or other organisations not normally under their command, which encompasses their responsibility for implementing orders or directives.' This defines a narrow scope, with command encompassing leadership tasks and control focused on execution.

At either end of the C2 spectrum, the differences between model approaches, those which are highly centralised and those which are more decentralised, can be thought about in terms of decision-rights, vested from an authority. Centralised models accumulate authorities and permissions in a few or a single individual, granting them the ability and power to make decisions. Conversely, decentralised models distribute authority and permissions to achieve perfect decision-making equality. With limited institutional-level examples at the bounds of each approach, the predominant application observed is a hybrid approach, borrowing features from each for a given context (Builder et al., 1999). As information and data become more readily available through the digitisation of means and methods of warfare, the pendulum is shifting towards more decentralised approaches to address the increased scale of information.

One decentralised approach is mission command, a philosophy of command where leaders provide intent[8], set the risk threshold, and define desired outcomes. Trust is placed in subordinates to execute with disciplined initiative, make local decisions and act autonomously to achieve mission objectives.

The concept of mission command is not new, with the practice recorded in some German units across World War I and World War II (Beavis, 2025). Its origins date back further, some 150 years, to the Prussian Army (Bunker, 2020), establishing the concept of *Auftragstaktik* (mission-type tactics). Mission command can be described through four functions, including: establishing intent; determining roles, responsibilities, and relationships; establishing rules and constraints; and monitoring and assessing the situation and progress (Alberts and Hayes, 2006). Mission command, by design, delegates responsibility through trust in leadership at all levels to make complex and time-sensitive decisions for combat advantage (Corn and Smotherman, 2025).

Many militaries espouse a mission command approach to C2, including the United States, the United Kingdom, Australia, New Zealand, and Canada, that build on early Prussian practices. The North Atlantic Treaty Organisation (NATO) has established mission command as the overarching alliance command philosophy (NATO, 2022). However, as Sjøgren and Nilsson note, 'while most Western militaries claim to adhere to the principles of mission command, there remain considerable differences in its implementation' (Sjøgren and Nilsson, 2025).

The increased integration of autonomous and AI-enabled capabilities into military C2 calls into question the applicability of all mission command principles to machine capabilities and functions. The specifics of these principles in machines are not considered in isolation; rather, they are considered in the context of how machines are integrated in C2 operations relative to their human counterparts.

As machine capabilities continue to evolve along with the culture and familiarity of these systems, the nature of their roles within C2 structures are shifting from tools to teammates (Neads et al., 2021). The applicability of the underlying C2 principles for machines are shaped by this dynamic. Table 1 outlines principles of military mission command amongst Western militaries.  The last column of Table 1 outlines the machine's ability to meet military mission command principles. A number of these principles are categorised as 'Yes and No' for machine applicability, highlighting the significance of context in military C2: the context of machines' capabilities and functions, the context of the role machines play in C2 and the context of those roles in relation to human roles and responsibilities. We will revisit these assessments through our discussion section.

---

[8] The Commander's intent is introduced and defined in the paragraphs following, within this section.

In this work, the key principles selected are highlighted in Table 1 and broadly include variations of trust, understanding, the Commander's intent, orders, initiative, and risk acceptance. **Error! Not a valid bookmark self-reference.**

We describe each of these selected principles through an Australian Defence Force (ADF) lens, principally as the authors originate from Australia. The ADF identifies that mutual trust "is the result of shared confidence across an organisation," and notes that it depends on competence in the context of an assigned task. In practice, this is a bi-directional relationship between a Commander and subordinates to establish the ethical tone of the organisation (Australian Defence Force, 2024). Shared understanding is the responsibility of the Commander to "ensure that everyone in their organisation" understands the problem that must be solved. It is closely coupled with disciplined initiative, for which it is a core requirement to provide subordinates a safe and trusted environment.

Commander's intent is a necessary condition for mission command, without which "it is not possible to exercise mission command" (Australian Defence Force, 2024). The Commander's intent is a "clear and concise expression of the operation's purpose, method and end state," providing guidance to enable subordinate action in lieu of further mission orders. The Commander's intent takes on many forms, although it is often characterised as a "concise description of how the commander intends to achieve the end state," including constraints, limitations, assumptions and resources.

Mission-style orders taken from the perspective of mission command focus on the function of the orders, rather than the form and style in which they are communicated. They are an essential aspect of mission command as the method to share understanding for the mission, applying the central principle of 'what not how' (Australian Defence Force, 2024).

In contrast to principles that outline requirements or conditions, disciplined initiative is described as a "duty to exercise initiative to achieve the desired end state." Tightly coupled to a manoeuvrist approach, disciplined initiative by subordinates offers Commanders an opportunity to seize the initiative or conduct through non-typical actions that may otherwise not be possible. Australian Defence Force doctrine notes that initiative is not an independent activity, necessitating trust between commanders and subordinates for success. (Australian Defence Force, 2024).

As our final selected principle, risk acceptance, means Commanders must consider the risk of action or inaction and manage accordingly to drive decisive action. It is a necessary principle that subordinates develop trust in their Commanders and that Commanders develop trust in their subordinates. It also sets the tone for how ambitious a subordinate may be during a mission (Australian Defence Force, 2024).

*Table 1: Principles of military mission command amongst Western militaries. Highlighted principles are those most frequently identified and are selected in this work as the basis of the analysis.*

| **Principles** | **NATO** (NATO, 2022) | **AU Army** (Australian Defence Force, 2024) | **US Army** (Bunker, 2020) | **UK Army** (Risso, 2024) | **NZ Army** (New Zealand Defence Force, 2017) | **CA Army** (Wadsworth, 2017) | **Machine Applicability** (Bunker, 2020) |
|---|---|---|---|---|---|---|---|
| Mutual Trust | X | X | X | X | | X | No |
| (Shared)(Mutual) Understanding | X | X | X | X | | X | Yes and No |
| Commander's Intent | | X | X | X | X | X | Yes and No |
| (Mission) Orders | | X | X | | | X | Yes and No |
| (Disciplined) Initiative | | X | X | | | X | Yes and No |
| Risk Acceptance | | X | X | | | X | Yes |
| Verification | | X | | | | | |
| Competence | | X | X | | | | Yes |
| Unity of Command | | | | X | | | |
| Freedom of Action | | | | X | | | |
| Unity of Effort | X | | | | | | |
| Timely and Effective Decision Making | X | | | | X | | |
| Decentralised Execution (Responsibilities) | X | | | | X | | |
| Commander's Determination | | | | | X | | |

## Directions in human-robot interaction

As a multi-disciplinary field, Human-Robot Interaction (HRI) focuses on the design and development of robotic systems for and/or by humans (Goodrich and Schultz, 2007), drawing upon fields ranging from robotics, AI, human factors, cognitive science, and sociology to create user-accepted systems that are both functional and safe.

Early work on HRI centred on aspects of robotic supervisory control, with the now-defunct master-slave paradigm dominant across the literature in the field. In addition to human supervisory control or teleoperation, HRI research has grown to include areas such as automated machine control, for instance, autonomous vehicles with human passengers, and human-robot social interactions, such as home assistance and entertainment (Sheridan, 2016b).

HRI has continued to evolve towards more flexible, team-based relationships. This is prominent within the interfaces and interaction modalities for HRI, moving from more static screens and joysticks to incorporate more human-based interaction methods, such as voice commands and gestures. The growth of multi-modal interfaces claims to offset cognitive load increases, enabling richer and more nuanced communication for HRI tasks (Walker et al., 2018).

Related to the concept of HRI are adjacent and often overlapping teaming fields, consisting of human-AI teaming (HAIT or HAT), human-machine teaming (HMT)[9] and human-swarm teaming (HST). In this work, we include HRI work that intersects with HAT, HMT or HST within the scope of the paper.

One of the most studied factors in HRI is trust, which is an important mediator for effective HRI. (Assaad and Boshuijzen-van Burken, 2023) define trust as 'confidence in the reliability of [a system] when used in the intended operation of use.' Trust is a dynamic attribute that calibrates over time, subject to system performance, reliability and predictability. Meta-analyses indicate that robot performance and attributes dominate trust formation, with human and environmental factors secondary in impact (Hancock et al., 2011). A key outcome in HRI is to develop an appropriate level of calibrated trust, mitigating the pitfalls of over-trust and under-trust. Over-trust involves overconfidence in a machine's capabilities, which may lead to human operators not detecting errors, malfunctions or incorrect outputs (Assaad, 2024). Under-trust manifests where an operator fails to utilise the full capabilities of the system, negating benefits and often leading to an increase in their own workload (Ruff et al., 2002).

Transparency is commonly posited as a means of practical trust calibration (for example in (Galliott and Wyatt, 2022) and (Zhang et al., 2021)). One example includes the Situation Awareness-based Agent Transparency (SAT) model (Chen et al., 2014), based on the theory of Situation Awareness (SA) (Endsley, 1995). This model supports operator SA in settings with human and non-human intelligent agents. Transparency and accountability frameworks building on the fundamental principles of SAT are emerging, framing transparency as a multi-dimensional construct to support trust and assurance in teaming activities (Hepworth et al., 2020). Other factors that were flagged in the literature as affecting trust in teaming activities include reliability (Lushenko and Sparrow, 2024) and the operational context of deployment (Galliott and Wyatt, 2022).

The concept of adapting existing human-centric lexicons for HRI purposes is not new. Adjacent to transparency research are emerging lexicons and formal languages for teaming with robotic and AI-enabled systems. For example, Abbass et al. propose JSwarm (Abbass et al., 2022). As an efficient communication language, it is inspired by the Australian Aboriginal Language Jingulu; and is intended as a language for guidance and control in HST activities. JSwarm leverages the underlying lexical structure of Jingulu to increase teaming efficiency, providing contextual relevance to the task (mission) at hand, while remaining computationally efficient and maintaining semantic equivalence with direct syntactic mapping. These desirable features provide an initial interface between humans and robotic agents.

[9] HMT is often synonymous with the concept of Manned-Unmanned Teaming (MUM-T).

HRI research today is moving towards genuine teaming constructs, departing from classic tool-based perspectives. This work considers robotic teammates not as instruments but as partners and investigates the conditions necessary for effective teaming in HRI in the context of military C2. Interestingly, Lushenko and Sparrow found that US military cadets were more willing to deploy in a HMT under a "centaur" paradigm (where humans directed robotic teammates) than under a "minotaur" paradigm (where the machine gave direction) (Lushenko and Sparrow, 2024). An important developmental area is the emergence of shared mental models, in which both human and non-human teammates form a common understanding of the mission, individual and collective tasks, roles and outcomes (Cannon-Bowers et al., 1993). One such model is the Ontology for Generalised Multi-Agent Teaming (Onto4MAT), which enables bi-directional information flow and grounding meaning for intelligent human and non-human agents (Hepworth et al., 2022). It should be noted that the machine-understanding concept does not imply human traits like reasoning, nuance, or subjectivity. Rather, it points to the more deterministic and binary nature of machine computation (Zhang et al., 2019).

## HRI in the military domain

HRI in the military is a growing area of research, with applications to both existing and emerging technologies. While the predominant employment of robotic systems within the military focuses on the remote control of uncrewed systems (UxS), advances in both AI and robotics are opening new opportunities to perform more advanced roles previously once only completed by human soldiers (Jentsch, 2016). While research and development continue to advance, a dominant approach that translates theory to practice has yet to emerge to address the limitations of contemporary systems.

With significant research and practical opportunities, HRI in the military has the potential to impact 'the practice of mission command in radically different ways' (Neads et al., 2021). Neads et al. signal the opportunity to evolve mission command for settings with non-human, robotic agents, highlighting the 'existing culture of mission command is seen as the gateway to future models of C2, in which deeper networking, autonomous systems and [HMT] will enable greater levels of self-synchronising tactical activity' (Neads et al., 2021).

Recent work has evaluated the applicability of mission command in HRI settings. The concept of mutual trust is defined by a shared confidence, based on demonstrated reliability and competence to perform all tasks (Australian Defence Force, 2024; Bunker, 2020). While unlikely to be applicable to machines (Bunker, 2020), a Commander may develop *confidence* in the behaviours and actions of a machine; the bi-directional requirement for shared experiences and common belief as mutual trust is not reciprocal in this sense from the machine to the human. The SAT model aligns directly with mission command's requirement for shared understanding and disciplined initiative.

Understanding, typically expressed as shared or mutual understanding, is the idea that a common frame of the problem, the operational environment, the mission's purpose, and its risks must be well understood by commanders and subordinates alike (Hepworth et al., 2022). While the idea of shared understanding and concept spaces is an active area of research (Hepworth et al., 2022), consensus has yet to be reached on its general applicability beyond technical interpretations to ethical and moral situations (Bunker, 2020). Nevertheless, such issues could be overcome with the careful design of human-machine interfaces (Neads et al., 2021).

The general concept of intent centres around a purpose, aim or goal (Crump, 2010). Commander's intent similarly focuses on a purpose or goal; however, that purpose is narrowed to the mission at hand. A commander's intent encompasses the purpose of the mission, the proposed method of achievement and the desired end state or outcome (Australian Defence Force, 2024). At its core, the idea of a Commander's intent provides the bounds on possible future states and actions of a machine teammate; however, the fog of war and deep uncertainty continuously present on the battlefield impact the effectiveness of machines and the underlying clarity of the context at hand (Bunker, 2020). JSwarm provides an initial bridge between doctrinal phrasing and the computational representations required for mission command. Recent research highlights the importance of considering Commander's intent throughout the lifecycle of AI to ensure clear command accountability aligned to specific AI use-cases (The Global Commission on Responsible Artificial Intelligence in the Military Domain, 2025).

Mission orders are the directions that articulate the result to be attained, although not the actions necessary to achieve them (Australian Defence Force, 2024). As with Commander's intent, mission orders have variable applicability to human-machine interactions – while simpler tasks and interactions have clear links to orders, more broadly scoped missions may have complex interactions, dependencies and implied actions, necessitating a broader context and understanding to assure adherence.

Disciplined initiative refers to independent action taken in achieving defined objectives, while remaining within the parameters established by the Commander's intent. The approach advocates for active problem identification, solution generation, and maintaining consistency with overarching mission goals (Australian Defence Force, 2024). As with many concepts central to mission command, the applicability of disciplined initiative is conditional on the context at hand, with more complex environments increasing the possibility of misinterpretation, acting beyond the acceptable command bounds.

Risk acceptance promotes the management of accepted risk, exercising initiative and acting decisively (Australian Defence Force, 2024). Diverging from most criteria, where the trade-space requires more clarity. The acceptance of risk is a well-suited task for machines, able to quantify risk and determine a course of action for execution. The ideas presented within the HST3 model proposed by (Hepworth et al., 2022) aligns with mission-command's requirement for shared understanding and disciplined initiative, providing a theoretical bridge for these concepts within military HRI. Notwithstanding, ethical, safety and social considerations inform guardrails for autonomy, assurance and governance mechanisms that MHC1 must embed across the lifecycle (design, testing, deployment, and in-operation control) (Assaad and Hepworth, 2025).

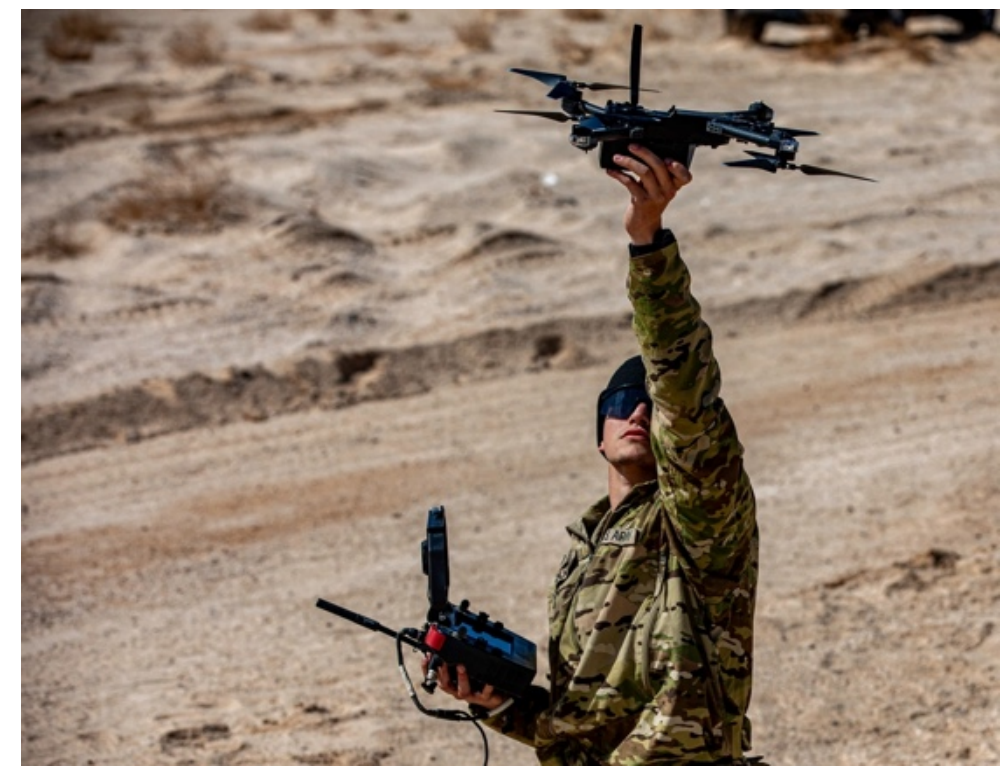
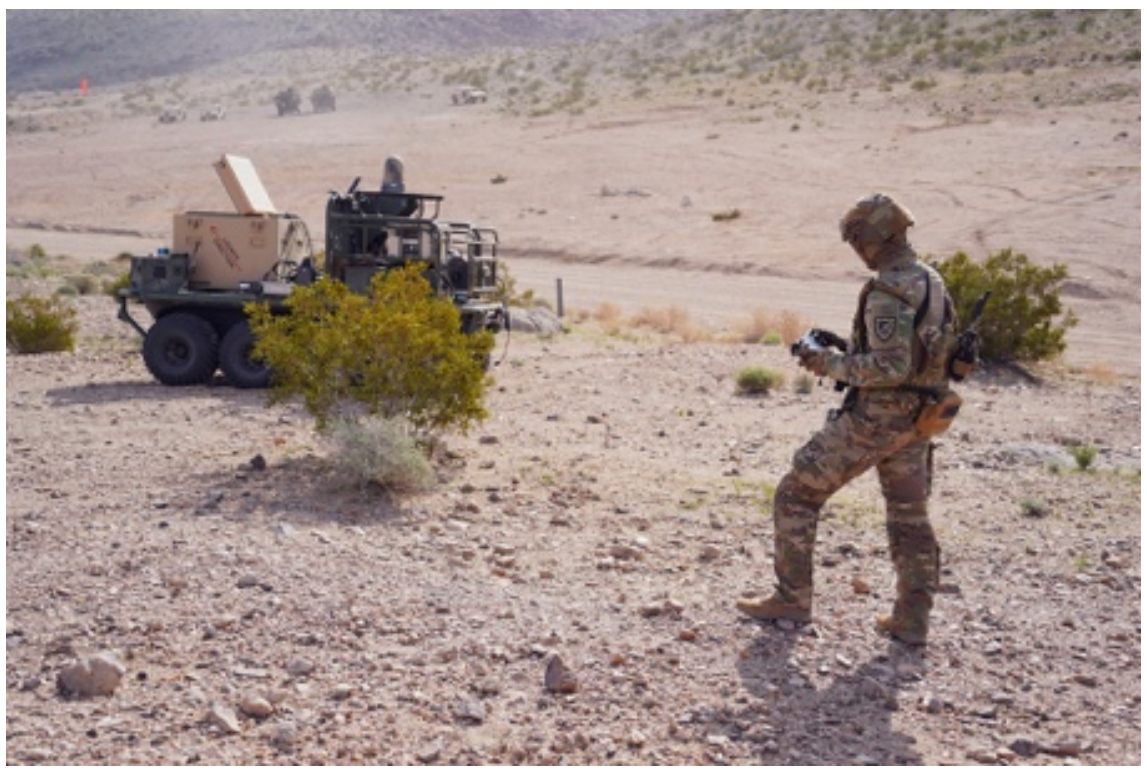
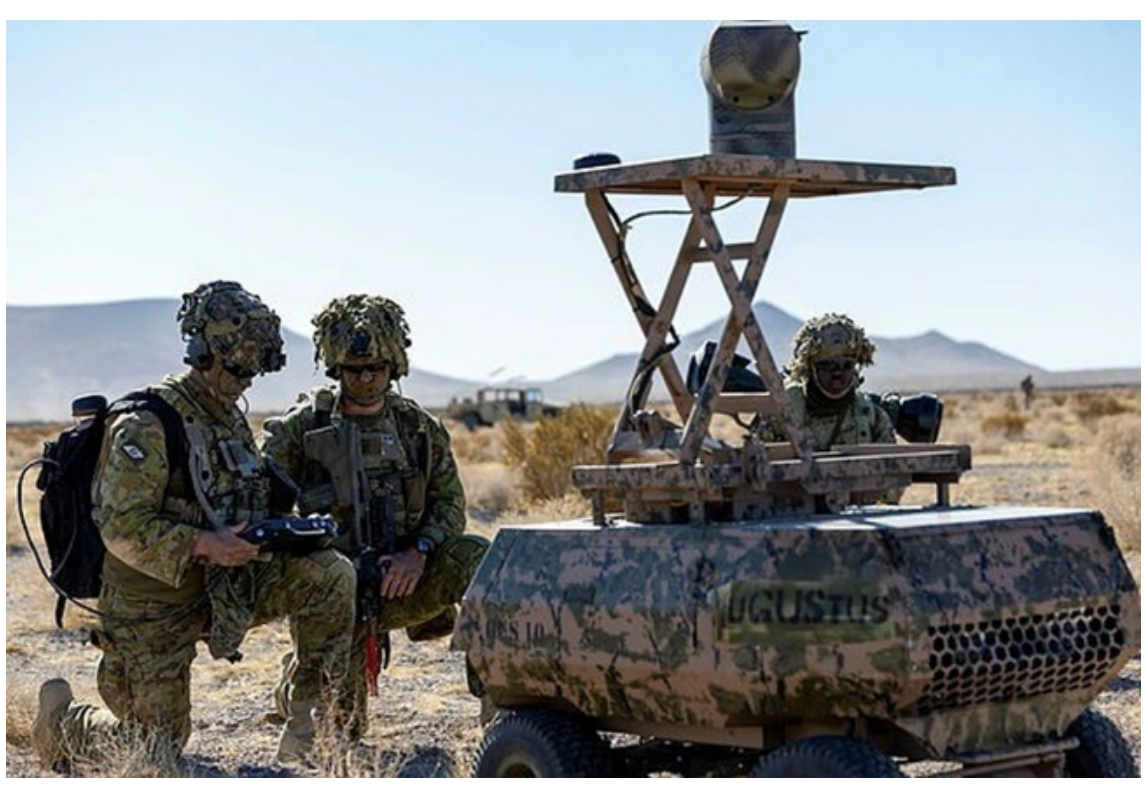

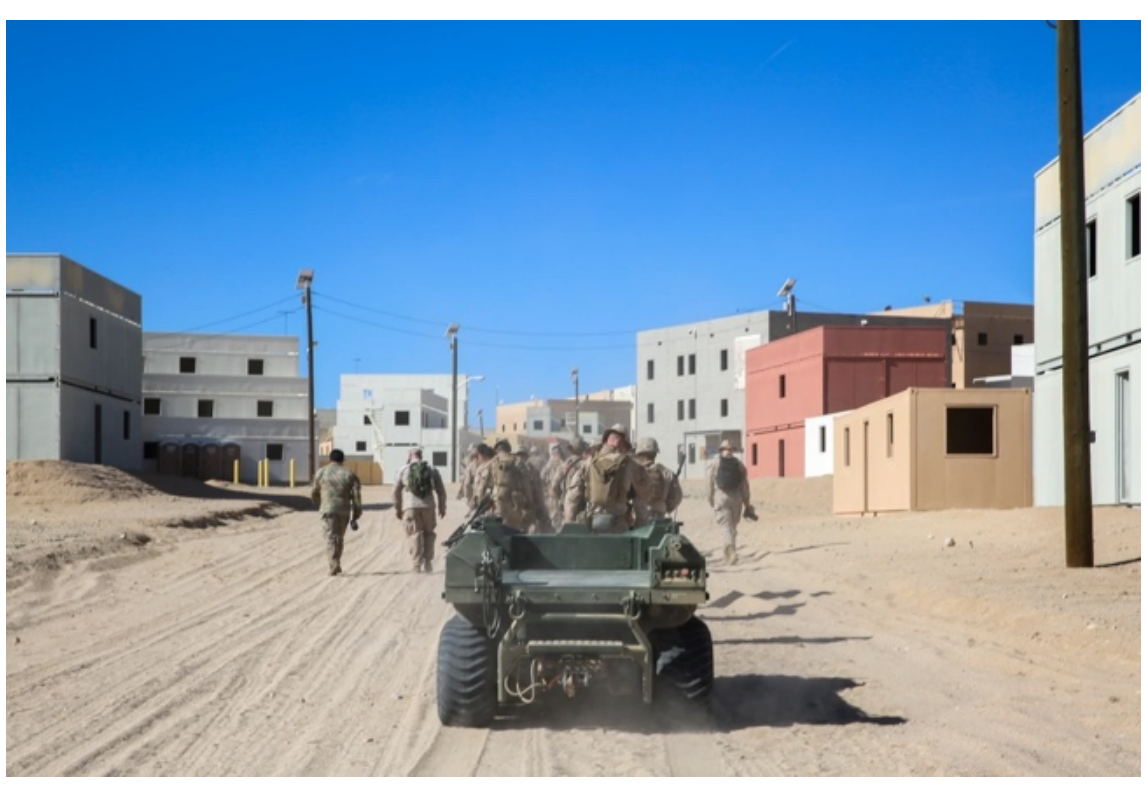

*Figure 2: Contemporary examples of military HRI applications (Top Left – SGT Matthew Wantroba/US Department of Defense; Top Right - SSG LaShic Patterson/US Department of Defense; Bottom Left – Jonathan Koester/US Department of Defense; Bottom Right - Kevin Ray Salvador/US Department of Defense)*

We propose that three key factors differentiate HRI in the military domain from other applications. These include the intent to harm, the presence of a cognitive adversary, and the impact on decision-making cycles.

The first factor, intent to harm, differentiates operations conducted in adversarial environments where military organisations are employed to use force. The employment of robotic teammates as an extension of force necessitates high levels of assurance and accountability for use. While the intent to harm not only increases ethical and legal challenges with the use of robotic and autonomous systems, it also further challenges the notion of MHC as a stress-promoting activity on the battlefield.

The second factor, the presence of a cognitive adversary, with an agent introduced who has an explicit objective to disrupt and ultimately defeat the military actions of an opponent. Military operations are both complex and contested, with such an adversary seeking to disrupt C2 functions through a range of kinetic and non-kinetic actions. Within the frame of HRI, control-based models may exhibit system brittleness across multiple input surfaces. For instance, reliance on persistent (continuous) and high-bandwidth network links, such as direct interventional states for an uncrewed system, is vulnerable to interference and degradation (Baltzer et al., 2025). As Baltzer et al. go on to note, MHC may not be possible where a continuous human is not possible (Baltzer et al., 2025). A command model, such as MHC1, assists in such situations by delegating decision-making authority for local execution, thereby increasing the system's resilience.

The third factor, decision-making cycles, seeks to gain advantage through faster sensing, deciding and acting when undertaking military actions, with models such as the Observe-Orient-Decide-Act (OODA) loop (Boyd, 1976) and Sense-Decide-Act (SDA) cycle (McVicar, 1984) developed to deliver on this promised potential. AI and autonomy, particularly in teaming systems, are seen as key enablers for decision-making acceleration. This creates a fundamental tension between the desire to leverage AI's speed and the need for human control within the decision cycle. MHC falls short here, as direct human control negates its advantages, generating a fundamental tension between the desire for oversight and the need to out-decide an adversary. AI and autonomy, particularly in teaming systems, are seen as key enablers for decision-making acceleration. This creates a fundamental tension between the desire to leverage the speed and the need for human control within the decision cycle. We suggest that, by leveraging the benefits of AI-enabled decision-making, a command relationship is a necessary condition. MHC1 offers a path to resolve this by moving the human's role from low-level execution control to high-level mission command, driving execution to the lowest possible level. Adoption of the MHC1 necessitates navigating a complex trade-off space, a concept spanning economic, social, and capability-based outcomes.

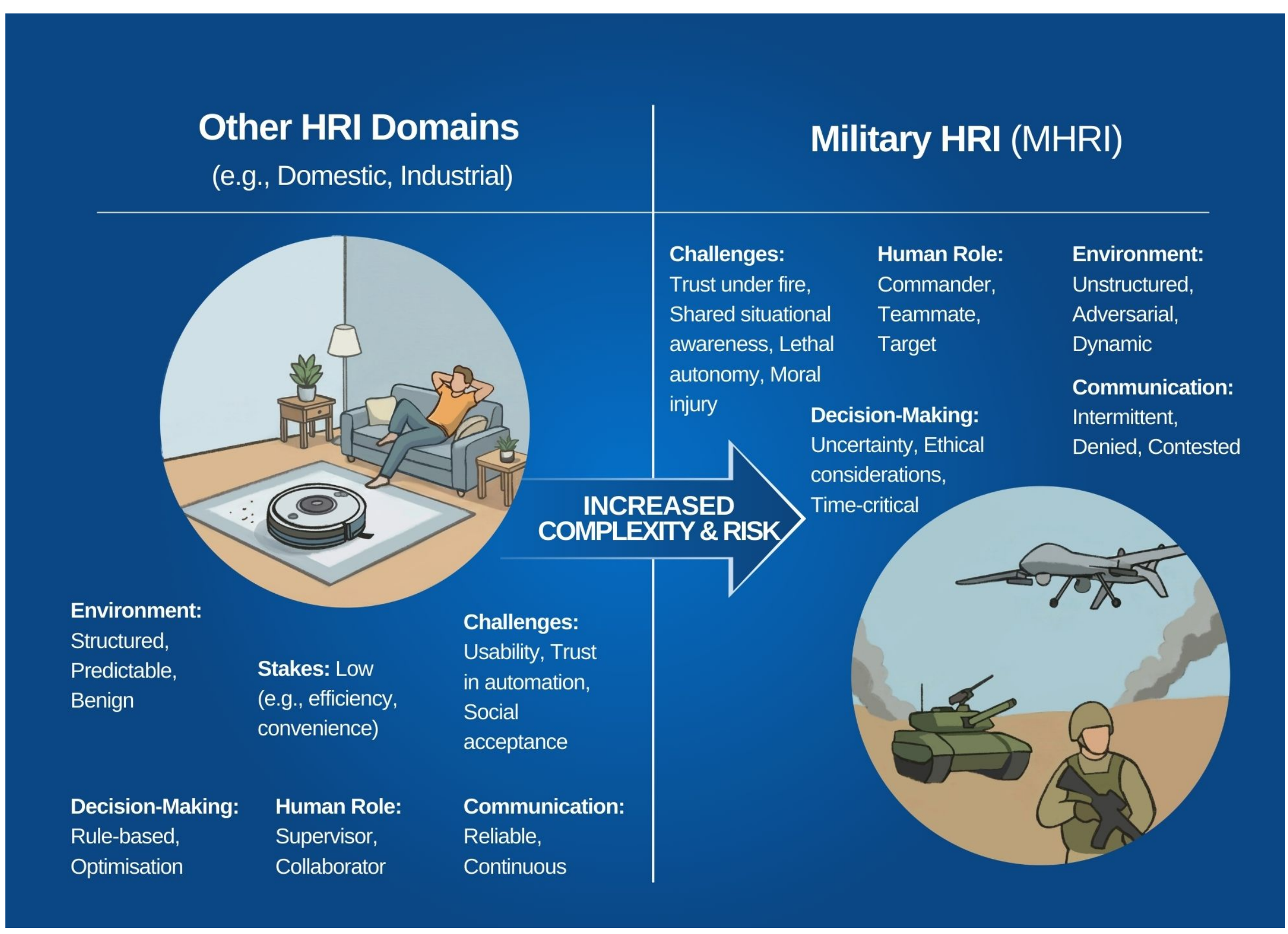

*Figure 3: Illustration highlighting key challenges differentiating HRI and MHRI.*

In the following sections, we further ground this work through definitions at the intersection of HRI and the military domain, explore an ethically focused vignette of military HRI, discuss the position of this work and propose Meaningful Human Command as a new approach to military HRI as our key contribution.

## Transitioning from Meaningful Human Control to Meaningful Human Command

A central premise of this work is to replace the discourse of *meaningful human control* with that of *meaningful human command*. We posit this is a necessary shift due to the inherent conflict between the concept of *control* and the principle of *autonomy* – not only from a perspective of autonomous machines, but also within human teams, where the term *control* is at odds with the notions of delegations and *subordinate commanders* with autonomy.

### Meaningful Human Control

The concept of Meaningful Human Control arose in response to the concern that fully autonomous weapon systems could operate across a wide geographic battlespace for extended periods without recourse to a human supervisor. Interestingly, this concern is based on primarily theorised understandings of the potential capability of fully autonomous weapon systems, rather than the reality of their current capabilities or military use-cases for their employment. While the technologies underpinning these systems have matured significantly over the years, it would be difficult to point to a current successfully deployed example of a fully autonomous weapon system. The notion of systems making decisions independent of human analysis or oversight has sustained debates around an "accountability gap" (Krishnan, 2009) or "responsibility gap" (Galliott, 2016). In reality, there exists a hierarchy of control and a chain of decision-making that leads to the point of deployment of such systems (Assaad and Williams, 2025). Therefore, a chain of responsibility is always present and cannot be obfuscated under the false pretence of an absence of human decision-making. The concept of retaining Meaningful Human Control as a precondition for ethical and legal use of autonomous weapon systems has nonetheless been a popularised point of inquiry for the international community.

Interestingly, despite its prominence in the discourse, there remains a lack of a concrete universal definition and commonly applicable technical standards; this is the case with many emerging concepts within this domain. In their papers (Ekelhof, 2019) and (Eklund, 2020) propose four common characteristics in extant definitions of meaningful human control. The first common element is the requirement that human operators be given sufficient training and experience with the weapon system such that they can make informed and accurate decisions about its use and the operational context, and the system is required to be designed in such a manner that it can provide that information clearly, concisely, and reliably. Secondly, the system must include a means for the human operator or supervisor to effectively intervene in its operation, either to modify its operating parameters, deactivate the system, or prevent the system from engaging a particular target. For such a capacity to be meaningful, the system must allow enough time for the operator to comprehend the incoming information and make decisions. There is a need to balance the time required for an operator to intervene effectively and the military necessity and advantages of acting. Thirdly, the system must be well-designed and extensively tested to ensure it operates in a predictable manner and does so with a high degree of reliability. Finally, the system must be integrated into an effective accountability framework that covers its life cycle from conceptualisation and design through deployment and decommissioning.

Drawing on this literature, we broadly define Meaningful Human Control as a supervisory relationship where a human operator has direct oversight and intervention over a system's actions. Notably, whilst control need not be at the level of teleoperation, there must be sufficient insight and interactivity for the human to have adequate awareness, the necessary decision space and a capacity to intervene. The challenge is how to achieve a sufficiently meaningful control capacity without limiting the system's autonomous capability. Adopting an instrumental view of human agency in the HRI may allow for sufficient control; however, it also minimises the practicality of autonomy in these systems. It seems paradoxical to pursue the deployment of autonomous systems only to then implement them with limited capacity to meet human control requirements. The notion of meaningful human control has consequently surfaced a negative tension between the argued benefits of autonomy and an antipathy for relinquishing human control. Meaningful human command presents an opportunity for alleviating this tension.

## Meaningful Human Command

Meaning human command reflects an organisational relationship founded within the principles of Mission Command, where the machine operates autonomously within the bounds of the human commander's intent. Reflective of the increasing capabilities of AI-enabled robotic systems, meaningful human command conceptualises machine actors as active agents in a heterogeneous team, their behaviours bounded and shaped by a set of controls. This is fundamentally distinct from a leader-to-agent monodirectional control relationship, bringing additional complexity and veracity to how HRI integrates into military systems. In practical terms, this is a human-machine command concept where authority is delegated through the Commander's Intent, empowering a subordinate (human or machine) to act autonomously with disciplined initiative to achieve the mission.

Current literature on meaningful human command is sparse, with Devitt's paper first introducing the term to the international relations lexicon (Devitt, 2023). Devitt introduces meaningful human command to address accountability limitations in the concept of meaningful human control, without necessitating a constant interventionist approach. The main idea presented employs advanced control directives as well-bounded actions that an autonomous system may conduct across a range of scenarios as a function for control, inspired by advanced care directives in the medical field.

The use of the meaningful human-command lexicon in the current paper differs from Devitt's. In this paper, we ground the concept within military mission command, where an autonomous system is delegated authority within the overarching intent of a human commander. In this way, this paper's conception of meaningful human command focuses on aligning the inclusion of non-human (machine) agents with the established policies and philosophy of mission command. Moreover, once the human Commander formulates the Commander's intent in a form that a machine can process and delegate a sub-mission to the machine, the machine can decompose the sub-mission and delegate each of the elements in the decomposition to other machines. The objective of this work is to bridge the gap between contemporary HRI mechanisms and how militaries execute C2 in practice.

A key distinction of the meaningful human-command approach is how it addresses the perceived accountability gap for autonomous and AI-enabled systems. While meaningful human control focuses on the capacity of the individual human to understand, engage with, terminate or override a robotic system, meaningful human-command adopts an approach to assurance and accountability based on team-based controls and norms, grounded within well-established military philosophy. By moving away from reliance on monodirectional or direct control at the individual human level, and towards a hybrid human-machine team-based system of controls, meaningful human command leverages established regulatory and doctrinal avenues to ensure accountability at the gestalt level of a military unit.

## Disciplined Initiative in the Human Robotic Context

Arguably, the most interesting trade-off space opened by a shift to meaningful human command is how to enable autonomous action whilst retaining meaningful guardrails around that machine's actions. In the mission command lexicon, the correct balance, where the agent is capable of independent action in pursuit of objectives whilst remaining within the barriers of the commander's intent, is termed "Disciplined Initiative". For human agents, this is conditional on the problem context, but influenced by training, team cohesion, common culture and trust. In the case of machine agents, the challenge of establishing meaningful guardrails based on a context-dependent commander's intent, whilst retaining sufficient self-direction and adaptability to capitalise on the agent's autonomous capabilities, is both technical and governance.

From a governance perspective, the challenge centres on how to develop and verify that the machine agent operates under the same contextual influences and norms as a human agent. This approach does not attempt to predict the actions of the machine agent, which may not be possible under every circumstance, given the non-deterministic nature of AI-enabled systems and the variety of potential decision scenarios in conflict. Instead, the aim is to shape the machine's decision-making process by designing the set of constraints and preferences to align the machine-agent's decision space with similar ethical, legal, normative and doctrinal factors that shape the Commander's decision space. This approach would require close coordination between designers and military end users, as well as extensive training for human-machine teams. Human commanders would need realistic training opportunities with machine agents to understand a given system's limitations and to gain experience in effective human-machine communication.

Among the unresolved questions associated with meaningful human command is the trade-off space arising from the impact of demonstrated high levels of autonomy on humans' tendency to anthropomorphise machine entities. For example, consider a scenario where an autonomous logistics ground vehicle is assigned to an infantry platoon but consistently prefers to stay close to the platoon sergeant, choosing to follow them unless given specific instructions otherwise. Such behaviour could easily be perceived to be the result of the platform bonding with the platoon sergeant or viewing them as its favourite human, an easily understood narrative that could further encourage humans to treat that platform like the platoon's military working dog. In actuality, a quirk in the algorithm's reward function has led it to weight familiar data more heavily when determining which beacon to follow upon deployment. Whilst this is a benign example, it highlights that as a system demonstrates greater autonomy within a system of controls, it will exhibit more non-deterministic behaviour than non-autonomous robotics, creating a richer ground for potentially problematic anthropomorphism.

## Many Roads Home: An Ethically Focused Vignette of Military HRI

In this section, we introduce a fictional vignette to anchor our discussion of military HRI and draw out the core ideas of the chapter, focusing on the multifaceted nature of cost illuminated by the unique characteristics of military HRI settings. This vignette provides a practical context to explore the complexities of mission command, human-AI teaming, and the associated trade-offs presented in this work. It the operation of an autonomous uncrewed ground vehicle (UGV) through distinct stages of a mission, each requiring dynamic decision-making by the autonomous system to adapt continuously to the evolving operational context. The capability detailed in this vignette is fictional. It is a theorised view of the potential development trajectory and employment of autonomous system capabilities.

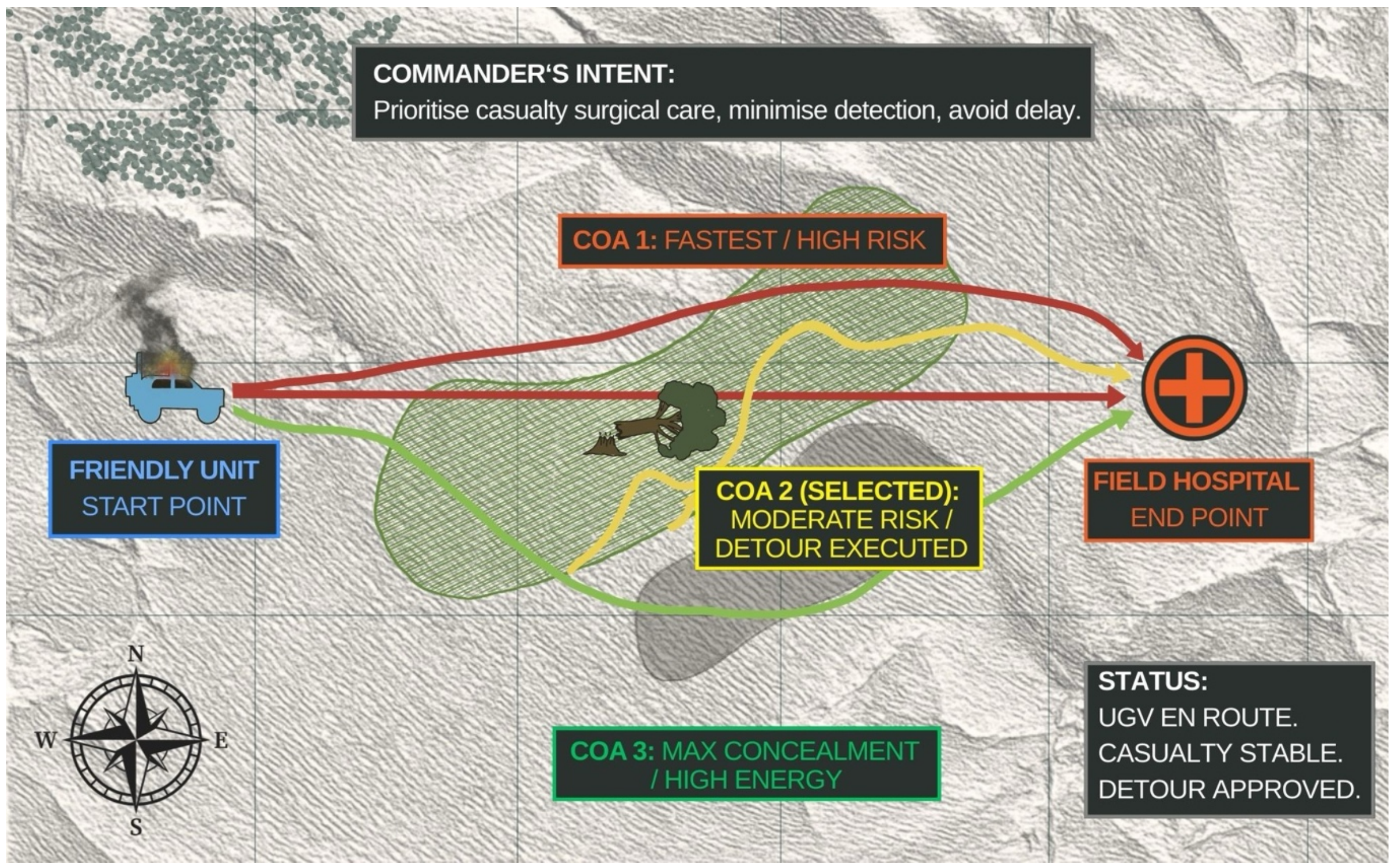

*Figure 4: Scenario illustration of key decision-making options throughout the vignette.*

## Part I: The Battle

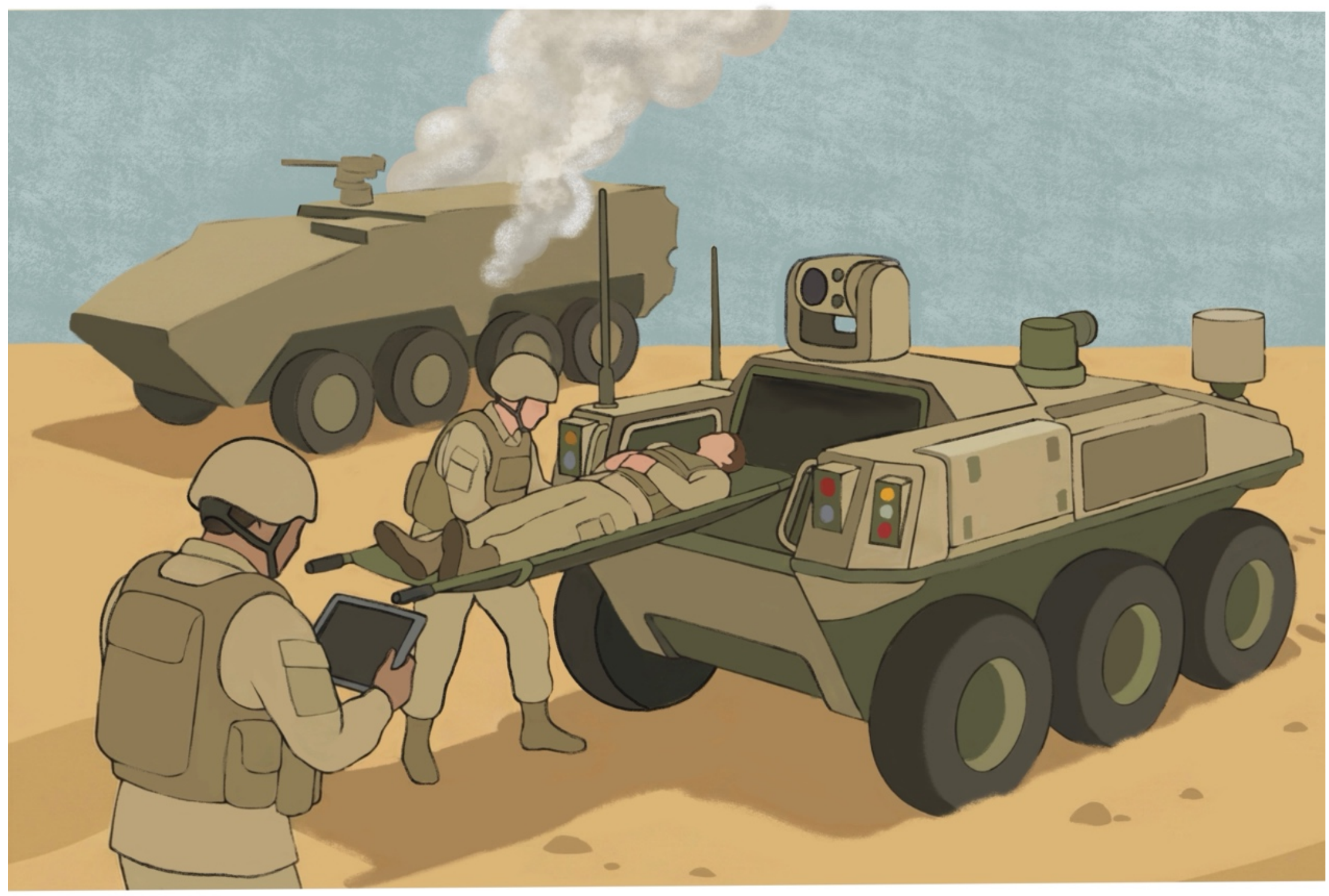

*Figure 5: A soldier is wounded during enemy contact, requiring the Commander to determine evacuation options to save the soldier's life.*

A shot rings out, followed quickly by sustained fire without warning. A precision loitering munition impacts an armoured vehicle near a team of dismounted soldiers – one soldier is injured and lies unconscious. The team conducts immediate first aid for the soldier behind the cover of the burning vehicle hulk. With the aid of stabilising the haemorrhage for evacuation, a priority medical request is sent – surgery is required immediately.

Shortly after the request for evacuation is sent, a response is received rejecting the forward deployment of airborne assets – surface-to-air threats in the area present an unacceptable risk to soldiers and systems. As options for evacuation quickly dissipate because of the ongoing battle, the Commander turns to one of the unit's AI-enabled Uncrewed Ground Vehicles (UGV) to undertake the casualty evacuation. More than a logistic task, the Commander must entrust the life of a human to one of their non-human teammates. As the UGV arrives, the combat medic prepares the wounded soldier for transport, loading the unconscious soldier into the UGV. Vital signs and telemetry from the wounded soldier begin streaming to both the medic's tablet and the UGV console in real time.

The Commander pauses and steps back momentarily to issue the UGV orders. In lieu of encoding these directly on the UGV's console as a specific route or other detailed information, the Commander provides high-level direction through natural language, giving the UGV a mission through a verbal order. The human commander outlines the evacuation priorities, key control measures and the desired end-state – delivering the soldier to surgical care. The Commander's approach for the interaction is embodied through the philosophy of mission command: retaining ultimate command and decision custodianship of the evacuation; *why* the mission is undertaken and *what* must be achieved. Execution of the mission is delegated to the UGV, with decision authority over *how* the mission is executed, and limits of authority outside of this context. By articulating the Commander's Intent, the Commander empowers local decision-making to the UGV. A mission command approach relies on shared understanding between the Commander and the UGV's AI-enabled operating system, which has been trained, evaluated and certified for such a mission in the months prior to the operation. While the principle of mutual trust does not apply in the same way with machines as it does with humans (Bunker, 2020), the Commander's approach demonstrates confidence in the UGV to act autonomously and interpret the human Commander's Intent appropriately, meeting the mission objectives within defined bounds.

## Part II: Evacuation

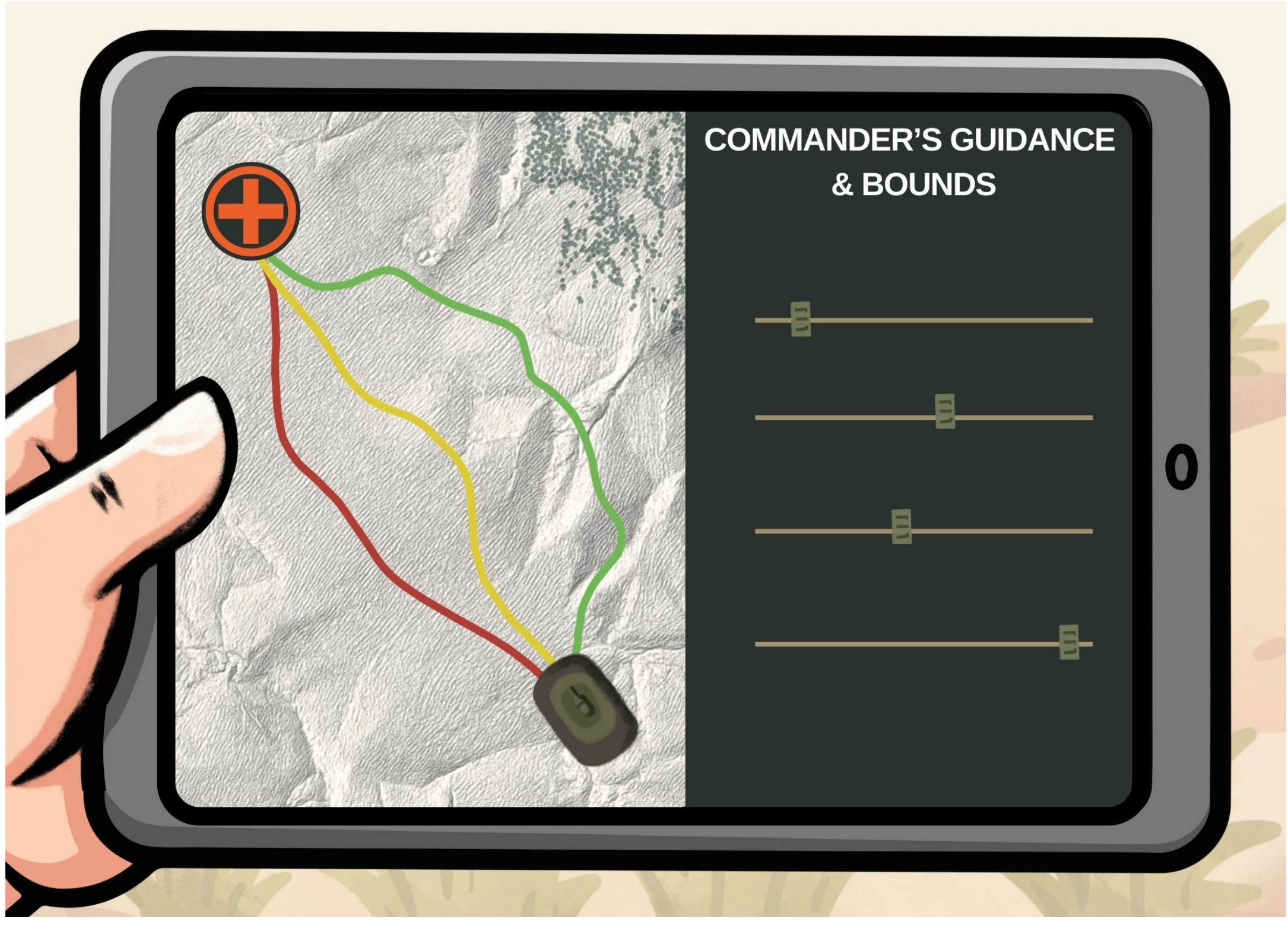

*Figure 6: The autonomous UGV provides routing options to the Commander that trade the welfare of the casualty with tactical necessity.*

As the Commander prepares mission orders, the UGV's console begins to present action options to the Commander, seeking guidance on their risk appetite, tolerance and preferences. The UGV presents three courses of action for the evaluation route, based on the Commander's most up-to-date direction and the overarching Commander's Intent. Each course of action balances discrete risk profiles, highlighting the breadth of risks and opportunities within the mission and the adaptiveness of the system, rather than a uniform behavioural response to the situation.

- ***Course of action one***: fastest route, highest risk of detection. This route prioritises the casualty's need for surgery as soon as possible, with a direct path over open terrain that enables faster transit. By focusing on delivering the casualty to the hospital quickly, the route becomes highly targetable and may unnecessarily expose the remainder of the unit to more concentrated enemy fire.
- ***Course of action two***: partially covered off-road route, moderate risk of detection. Concealed movement under tree canopy reduces the active drone threat and enables a clean departure from friendly forces. This route, however, often has debris and large objects blocking the path, which could make the path untenable, challenging mission progress and the UGV's all-terrain capabilities.
- ***Course of action three***: tactical route, low risk of detection. A highly concealed route that exploits the local terrain, minimising exposure to enemy forces; however, substantially impacting overall speed – the terrain is steep, narrow and energy-intensive, with the potential to exhaust the UGV's traversability and energy stores.

The Commander considers the merits of each course of action. Key indicators, such as mission completion time, probability of exposure and detection, energy consumption, probability of mechanical failure, and an estimate of mission complexity, are provided to the Commander. Reviewing the mission focus and complexity, the Commander determines that option three meets the least overall outcome, requiring frequent adaptation to the environment while maintaining visibility of the soldier's casualty state. The Commander notes that a simpler autonomy task, such as driving down a clear road, entails a lower cognitive load. In contrast, autonomy tasks, such as navigating obstacles under threats from an active enemy, increase the overall mission's complexity.

The Commander continues to adjust specific mission parameters, investigating key trade-offs to determine the preferred course of action. These trade-offs explore options such as the likely impact of speed on risk of detection and whether an increased sensor sensitivity will mitigate terrain uncertainty.

After the trade-off analysis, the Commander determines that course of action two offers the optimal balance: improved concealment to protect remaining friendly forces, while still meeting the time-criticality objective for the soldier. The Commander advises prioritising concealment over speed within the allowable mission time. The order is issued: *'Proceed with course of action two. Prioritise concealment over speed. If your detection exposure rises above nominal limits, report back immediately. If you deviate from the planned route, which results in more than a 5% increase in the total expected travel time, notify me with a recommendation.'* Notably, the Commander provides contingency actions for the UGV to govern future behaviour and its freedom to make autonomous decisions for exhibited behaviours as the situation evolves, identifying situations when human input is required for decision-making as a real-time reallocation of authority. The Commander provides guidance to the UGV, enabling it to exercise autonomous decision-making within approved bounds. This adaptation is relative to the mission context and local situation.

The mission orders clearly specify a Commander's Intent, priorities and constraints for the UGV to execute to transport the injured soldier safely, avoid detection and avoid substantial delays. The details of how this is conducted within the Commander's allocated bounds are now up to the UGV. As with briefing a human teammate, the Commander receives a back brief from the UGV confirming the mission, noting the selected route and highlighting key mission constraints. The Commander's direction necessitates that the UGV is ready (trained, equipped and prepared) and able (possesses sufficient capability) to complete the mission. Also highlighted is the Commander's willingness to accept the associated risks of undirected action afforded to the UGV without further shifts in authorisation. Explicit granular instructions are avoided, empowering the UGV to act in accordance with the Commander's Intent.

## Part III: The Mission

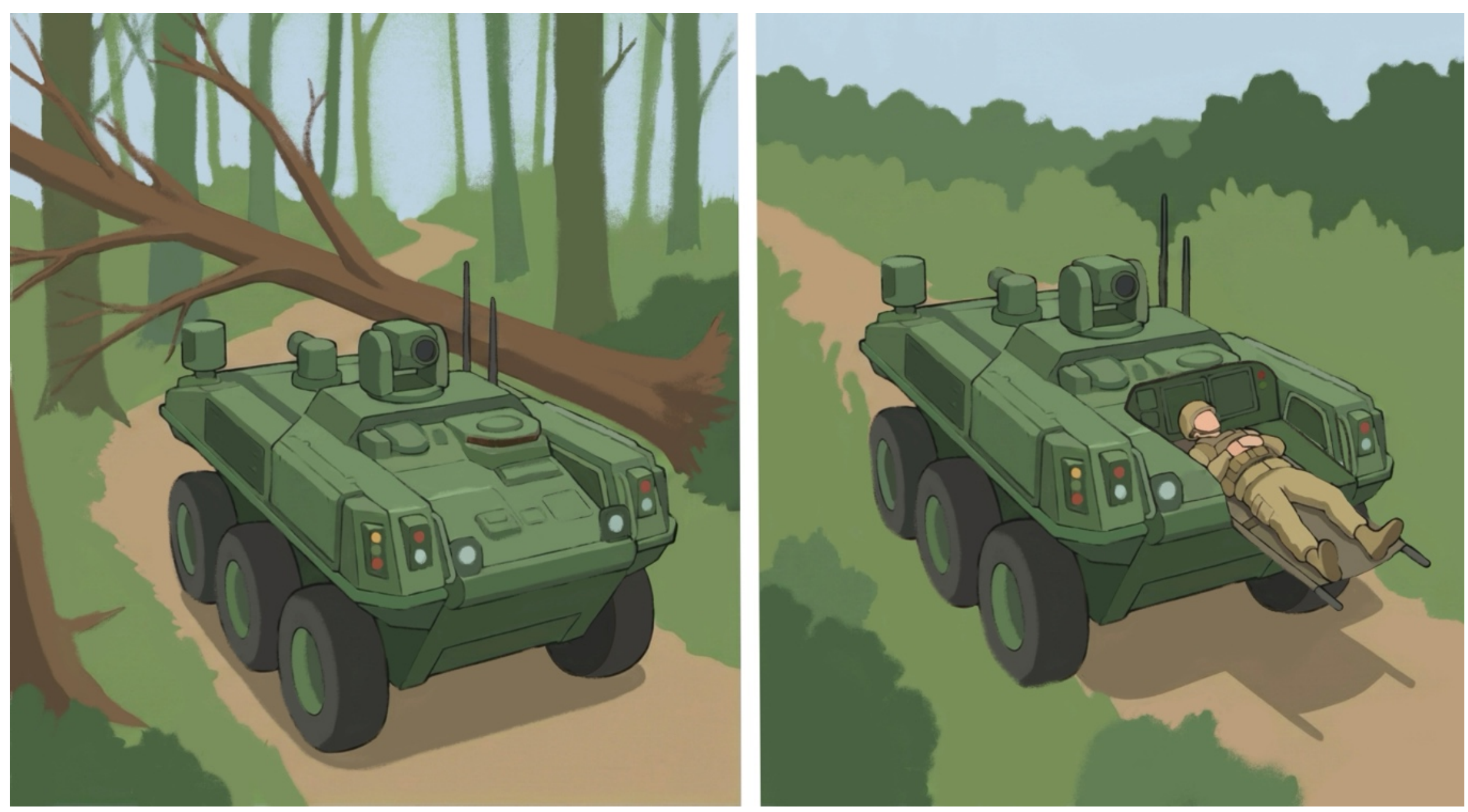

*Figure 7: During the mission, the autonomous UGV must trade-off different priorities and constraints to achieve the mission, reflecting an autonomous implementation of the mission command concept.*

The UGV sets off on the task, moving away from the unit as the battle continues. The soldiers are uneasy as they watch a colleague and friend, unconscious, being taken away by the UGV. They know, however, that a traditional evacuation would take more soldiers from combat, placing even more people in harm's way. Periodic updates are shared with the Commander as the mission continues, with the UGV's exhibited behaviour adapting dynamically to changes in the local environment and the evolving battle.

The UGV senses increased dust plumes from transiting the trail – more than anticipated, likely due to the recent drier-than-usual conditions. The UGV detects this change and reduces speed while options are developed, deciding to maintain the reduced speed to avoid exceeding the moderate risk profile. The adjustment is calculated to increase the total transit time by 2-3%, within the UGV's delegation limit. A short message is sent back to the Commander: 'reducing speed to maintain concealment with elevated signature'. The Commander acknowledges the situation change, concurring that the decision is in accordance with the mission Commander's Intent.

Now moving more cautiously through a dense section of the route, the UGV increases the sensitivity of its sensors, within the assured range to avoid unknown obstacles and the possibility of becoming immobile. Up ahead, the track is blocked due to a fallen tree – impassable due to the UGV's size within the time constraint. While alternative paths exist, they would deviate beyond the allowable limit set by the Commander, increasing the total transit time by 7%. A burst update is sent to the Commander: A *7% detour is required to meet the detection threshold and concealment bounds—increased risk to mechanical reliability and system task availability following the deviation. Casualty remains stable*. This marks the transition to a consultative mode for the UGV, adapting its decision-making level to meet the Commander's Intent risk threshold and seeking guidance for actions beyond the approved envelope.

Almost as soon as the message is sent, a warning is received: the message has failed to reach the Commander. Embedded redundancy measures within the UGV result in a second follow-up message: *Communications dead-zone– executing immediately to buy back deviation time*. The change in context resulting from the communication dark zone now necessitates the UGV to utilise greater degrees of freedom for decision-making, evaluate options against the Commander's Intent and relax directed mission constraints that may limit disciplined initiative to achieve the mission, demonstrating a failsafe-mode action at a higher level of autonomy. The Commander's Intent was clear: save the life, maintain cover and avoid delay. Partway through the deviation, the Commander receives the notification and formally confirms the deviation. The Commander notes that the UGV has exceeded its risk threshold, in service of the greater mission intent, responding: *'Detour approved –-continue mission.'* The detour, while originally unapproved, represents mission command in a human-machine teaming context akin to what would be expected within a homogeneous human-human team context, reaffirming the mission competence of the UGV to dynamically assess risk and allocate resources when the change in context exceeded the parameters of the original mission situation.

## Part IV: Mission Complete

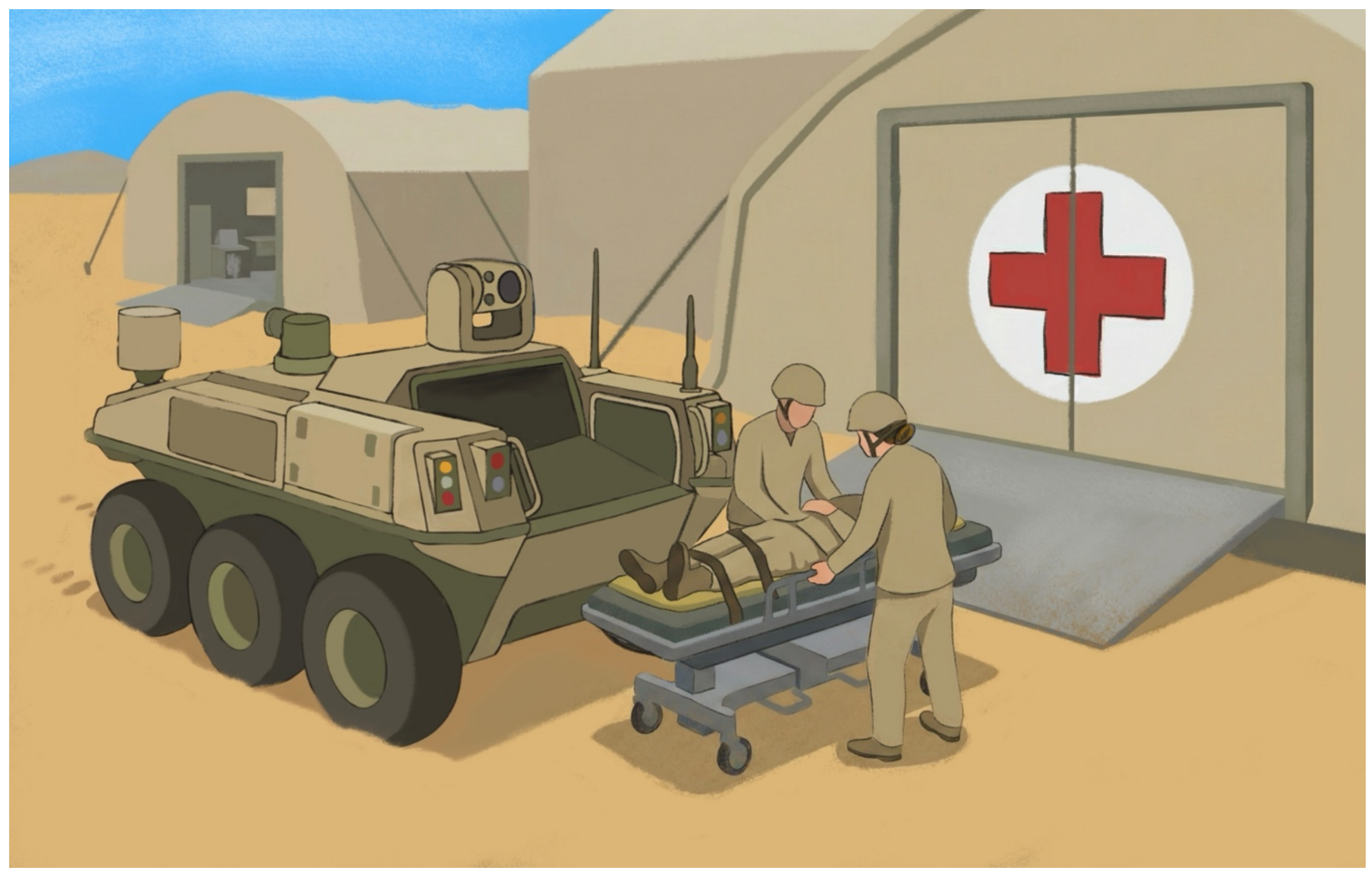

*Figure 8: The soldier casualty is off-loaded from the autonomous UGV, concluding the mission.*

The UGV approaches the field hospital and identifies the waiting medical teams. The medics guide the UGV into the surgical bay entrance, while medical telemetry begins streaming in the operating theatre, allowing the surgical team to prepare for the task ahead. The UGV evacuation log provides a summary of the initial trauma to the medical team. The UGV's mission to deliver the soldier to the hospital within the constraints provided is achieved.

Now transitioning into a higher decision-making capacity, the UGV enters a post-mission analysis mode and conducts an autonomous review of the mission. The post-mission analysis assists the UGV to optimise its own systems into the future, while also providing additional details to the unit and other evacuations – be they shepherded autonomously or by a human.

# Discussion

## Towards applying mission command in military HRI settings

Our philosophy through the vignette is one of *human command, AI-supported control*, where the Commander retains ultimate authority over the mission and its outcomes while delegating portions of the execution to AI-enabled systems. The key idea here is that MHC1 promotes human-centric decision-making, sharing control with AI-enabled systems when appropriate. By decoupling command (establishing mission parameters through the Commander's intent) from control (local execution), humans are supported by AI-enabled systems to perform tasks that may otherwise exceed their abilities or capacity during the mission. We will illustrate this perspective through the vignette in the previous section, with Table 2: Summary of agents within the vignette. and Table 3: Summary of key episodes within the vignette. summarising the key actions, agents and events across the scenario to support our discussion.

*Table 2: Summary of agents within the vignette.*

| Agent | Role |
|---|---|
| Commander | The Commander serves to define the mission intent, select and guide courses of action, establish risk tolerances and ultimately retain focus on the 'what' and 'why' of a mission. |
| Autonomous UGV | Responsible for executing the casualty evacuation mission, navigating local terrain, monitoring the casualty and evaluating local risk impacts to the overall mission. Focus on the 'how' of the mission. |
| Combat medic | Prepares the soldier for evacuation, monitors telemetry and communicates with the field hospital regarding the soldier's state prior to departure. |
| Wounded soldier | Soldier temporarily lacking independent decision-making abilities due to injury. |

*Table 3: Summary of key episodes within the vignette.*

| Episodes | Relevant agents | Description |
|---|---|---|
| Initiation | Commander | Determines that, given the tactical situation, evacuation should be conducted by an autonomous UGV. A key decision factor is the impact of different evacuation methods on the unit's warfighting capability and capacity. |
| Planning | Commander and Autonomous UGV | Commander establishes bounds and risk thresholds, and the autonomous UGV develops options based on the current situation and environmental |

| | | context, as well as its own capabilities and capacities. |
|---|---|---|
| Execution | Autonomous UGV | The autonomous UGV monitors the warfighting situation, local environmental context and casualty to dynamically make decisions that best achieve the Commander's Intent. |
| Transition | Autonomous UGV | The autonomous UGV facilitates handover of the casualty to the field hospital, subsequently initiating an autonomous post-mission review. |

We draw out the principles of mission command from the vignette to demonstrate their application to uncrewed systems. The Commander commences by issuing mission orders that specify the purpose of the actions, bounding the objectives of the autonomous system. In this situation, the mission is made clear. The commander articulates constraints through the risk posture and reporting triggers, delegating responsibility for local execution to the UGV. Aligned with doctrinal definitions, the Commander's intent is clear, supporting disciplined initiative and risk acceptance to achieve the directed end state. The point of interaction between the Commander and UGV is abstracted from continuous intervention (manual control) to the system of command.

The vignette demonstrates the value of mission command as a frame of reference for HRI in military settings, enabling Commanders to allocate risk to non-human agents where opportunity or payoff exists; in this example, to save a life and preserve combat power. This shows the importance of mission command in maximising the benefits of autonomous capabilities. By providing a clear Commander's Intent and delegating authority within bounds, the Commander was able to allocate outcome risk to a machine to save the soldier's life and keep the human, and possibly other, autonomous teammates employed in their primary functions.

The vignette describes four foundational functions from the C2 literature, often associated with mission command. These include the establishment of intent, allocation of roles and relationships, setting of rules and constraints, and monitoring of the execution  (Alberts and Hayes, 2006).

While the mission command principle of mutual trust is not applicable to machines (Bunker, 2020), the Commander develops *confidence* with the machine, the bi-directional requirement for shared experiences and common belief as mutual trust is not reciprocal in this sense from the machine to the human, although this does not prevent effective teaming within a mission command construct. Nonetheless, this does not negate the requirement for a system to operate within the situational context predictably (Helmer et al., 2024).

Emphasised throughout is the need for shared understanding, developed through regular communication, reporting and clear guidelines on the limits of authority that the UGV was able to pursue, employing a *trust but verify* approach to ensure accountability for mission outcomes. This approach builds on the foundation of a broader assurance pipeline, relying on the facilitation of a repeatable, real-time capacity for AI-enabled autonomous systems to provide transparent and explainable[10] outputs and behaviours to the commander (Helmer et al., 2024). While enabled to act autonomously with initiative, the UGV's outputs and decisions remain observable and verifiable, as with any human decisions made in a similar situation. Achieving shared understanding across teams of cognitively heterogeneous human and non-human agents is an open research challenge, necessitating bi-directional transparency and explainability (Hepworth et al., 2020). This marks a significant shift from one-way control interactions towards reciprocal, team-based relationships as a defining feature of contemporary HRI research.

While the machine cannot truly replicate trust, it is designed with the socio-economic and cultural perspectives of the originating nation and instantiated with those values. Whereas the mutual trust quotient may not be obtainable in this setting – the UGV isn't trusting the humans back – the outcome of the mission demonstrates that a functional one-way trust – humans trusting the machine's capabilities within defined bounds – can possibly be sufficient for the vignette's context, provided reliability and human oversight are maintained. These nuances are captured in the range of decision-making freedoms afforded to the machine and adaptive human-machine interactions within the vignette, with the UGV transitioning from direct supervision and manual control during planning, through to higher-order decision-making categories.

## Approaching meaningful human command

Emerging from the perceived accountability gap concept around autonomous weapon systems, MHC prioritises a supervisory relationship where a human operator maintains oversight via direct supervision and the capacity for intervention. This leads to a fundamental tension among the purpose of an autonomous system, the agency afforded to it, and the need for instantaneous oversight. In adversarial, time-compressed and communications-degraded situations, strict interpretation of MHC assumptions conflicts with the conduct of C2 through mission command. Continual operator observability and the need for immediate intervention increase an adversary's attack vectors, exploiting systems through methods such as electronic warfare. Moreover, the control-centric paradigm is misaligned with how militaries practically apportion responsibility, both legally and institutionally, through command.

---

[10] Effective command requires a level of mutual understanding, which is complex in the case of AI-enabled systems. For an overview of the delta between explainable AI and understandable AI see: Abbass, H., Crockett, K., Garibaldi, J., Gegov, A., Kaymak, U., & Sousa, J. M. C. (2024). From explainable artificial intelligence (xAI) to understandable artificial intelligence (uAI). *IEEE Transactions on Artificial Intelligence*, *5*(9), 4310-4314. (Abbass et al., 2024)

MHC1 attempts to reframe this situation and resolve the tension through a deliberate transition from supervision to technical and procedural oversight. Within the MHC1 construct, local control is delegated within certified and bounded behaviour and performance envelopes. This system-of-systems approach supports team-based system-of-controls, already established within existing military structures. When viewed through this lens, existing assessments of mission command applicability may be more aligned than previously considered. The vignette provides a basis to re-evaluate the applicability of these principles in contrast to the assessment of select principles previously analysed (Bunker, 2020). Table 2 describes our assessment, noting that the competence principle has been withdrawn from this contrastive assessment of (Bunker, 2020), with principles identified as most frequent across different militaries selected in this work.

*Table 4: Assessment of mission command applicability to uncrewed systems, contrasted to (Bunker, 2020).*

| **Principles** | **Machine Applicability** (Bunker, 2020) | **Revised assessment of machine applicability (from this work)** |
| --- | --- | --- |
| Mutual Trust | No | No |
| (Shared)(Mutual) Understanding | Yes and No | Yes (functional confidence) |
| Commander's Intent | Yes and No | Yes |
| (Mission) Orders | Yes and No | Yes |
| (Disciplined) Initiative | Yes and No | Yes (functionally bounded) |
| Risk Acceptance | Yes | Yes |

For mutual trust, the relationship remains asymmetric as the machine in the vignette cannot trust a human. However, functional confidence in the system can be developed by the Commander to enable delegation within defined bounds, serving a specific purpose within mission command.

With shared understanding, the machine cannot achieve the same level of subjective human understanding; however, it can develop a functional model of the mission, objectives, constraints and Commander's intent. This, in turn, is achieved through the execution of mission command within the vignette.

The autonomous UGV within the vignette is designed to operate within the bounds of the Commander's intent. This provides the high-level objectives and constraints to guide the machine optimisation and decision-making processes. Mission orders are well-suited for autonomous systems, providing key constraints and objective states without specifying the optimal execution method.

Disciplined initiative is a core tenet of mission command, described throughout the vignette. The UGV exercises disciplined initiative to locally problem-solve the path deviation request. While not explicitly programmed for the specific eventuality, the sequence of higher-order actions selected from the set of atomic actions was derived from the Commander's intent. Mission parameters, risk thresholds, and failsafe protocols shape the actions of the autonomous system.

Machines are well-suited to risk acceptance, both for quantifying and assessing risk's consequences. When presented with clear options and associated risk profiles, based on features such as time, tactical exposure, and mechanical failure, the Commander was enabled to make informed decisions on the course of action and delegate management of this to the UGV.

Translating a real-world mission based on the theoretical vignette to practice necessitates more than a philosophical fit with the concept of mission command. It requires a machine-actionable encoding for the series of atomic actions, higher-order activities, constraints and the Commander's intent, for which the autonomous system is expected to abide. Existing teaming constructs, such as Onto4MAT, could drive this encoding (Hepworth et al., 2022), where mission command defines who decides, for or about what, and with what accountability. The approach described within Onto4MAT defines how the decisions, roles, effects, constraints and justifications are represented to enable autonomous agents to execute the mission.

## Key Considerations

We offer five considerations for why MHC1 could provide organising principles to integrate HRI more deeply into military HRI. We argue that MHC1 provides a framework for translating key tenets of Mission Command to machine teammates, without eroding human authority and responsibility. The framework sets a foundation for clarifying decision rights, requirements and interfaces for military HRI.

Consideration 1: MHC1 offers a way to responsibly scale AI in military operations through a human-centric approach, while maintaining fundamental human oversight aligned with established military C2 paradigms. Existing mission command principles provide the basis for the success of MHC1, with human teams well-versed in communicating mission purpose and end state. While not all principles translate directly from human-human teams to human-machine teams, bounded applications remain possible within the MHC1 framework.

Consideration 2: MHC1 espouses a philosophy of *human command, AI-supported control*, ensuring accountability is transparently retained by humans throughout operational use within the mission command framework. In contrast to MHC, this generalises the concept of continuous and discontinuous control by allowing actions to be generated within a defined and assured framework to deliver on the Commander's intent.

Consideration 3: MHC1 enables both military practitioners and HRI researchers to develop an aligned and shared understanding to advance HRI outcomes within the military domain. As a mature framework for military operations, mission command is a human-centric approach that affords both military practitioners and HRI researchers the ability to develop a shared understanding of the Commander's intent, constraints and decision rights between humans and machine teammates. Our reframing of 'meaningful' through a lens of outcomes-based rather than actions-based provides a basis for closer alignment between key stakeholders.

Consideration 4: MHC1 will require policy, procedural (operational) and technical coherence during implementation to ensure RAI adoption. This requirement aligns across all systems of enabling infrastructure, with a clear commander's intent, authority and accountability structure to support governance, use and system development. This establishes a shared understanding for delegation, escalation and risk, and therefore ensures consistency from design through to development and operations. MHC1 couples decision-making, training and enabling infrastructure with feedback to update doctrine, procedural use and design to align with the Commander's risk tolerance.

Consideration 5: MHC1 generalises beyond autonomous weapon systems to all AI-enabled systems in the military domain, offering a path from existing MHC implementations to broader MHC1 approaches, as a viable basis for all AI-enabled military functions.

## Future HRI Work

Table 1 outlined mission command principles and presented Bunker's perspective on whether a machine can adhere to some of them. Our list exceeded the principles that Bunker studied. In Table 4, we updated Bunker's opinion based on lessons learnt from recent advances in artificial intelligence and the review we conducted for this chapter. Some elements in Table 1 were not discussed in the machine context: Verification,

Competence, Unity of Command, Freedom of Action, Unity of Effort, Timely and Effective Decision Making, Decentralised Execution (Responsibilities), and Commander's Determination. Each element in Table 1, whether studied by Bunker or us, suggests research opportunities and open questions for HRI and MHRI. Our discussion here is not intended to be comprehensive; instead, we will provide only examples of these opportunities.

To date, mission command frameworks have been human-centric. Transforming these frameworks to a human-machine or machine-machine setting presents technical challenges; the most prominent of which are metrics and modelling. For example, there is a significant HRI literature on shared understanding. However, this literature measures shared understanding only from the human side (see: (Chiou et al., 2021), (Bruemmer et al., 2005) and (Riley et al., 2016)). So far, the focus has been on measuring humans' understanding of the machine using questionnaires and scales and identifying which information to display to foster this shared understanding. Unfortunately, these approaches come with significant limitations, including the fact that shared understanding occurs in dynamic situations where the form of questions and the information presented to humans vary over time, and machines' understanding is neither measured, formalised, nor even defined. A large literature exists on 'machine understanding' (see, for example, (Jebari and Greenberg, 2026)). However, most of it discusses the philosophy of a machine that understands, assesses understanding in a conversational-like dialogue with a Large Language Model (LLM), or conflates understanding with interpretability, explainability, and visualisation. Recently, some researchers explored the use of ontologies as a bridge (for instance, (M et al., 2025)), which is a research direction with merit that can leverage recent work in HRI (see: (Hepworth et al., 2024)). Clearly, there remain significant opportunities in this research area to objectively model and measure 'understanding'.

A few mission command principles have been the subject of inquiry in computer science for decades: verification, competence, and timely and effective decision-making. Technological solutions exist to address them, but complexity challenges remain when scaling up their implementation. Other mission command principles create new research directions in HRI, ranging from concept conceptualisation and definition to metrics and modelling: Unity of Command, Freedom of Action, Unity of Effort, Decentralised Execution (Responsibilities), and Commander's Determination. Some existing HRI literature can contribute to some of these principles. For example, related work in trusted autonomy, including definitions and metrics of autonomy, and modelling research on how an agent's constraints and objectives shape their autonomy and behaviour, can contribute to the freedom of action principle. Other principles, such as unity of effort and decentralised execution, are nearly non-existent topics of from the HRI literature and require more foundational work.

In summary, the mission command principles and the meaningful human command requirements shape significant research opportunities for the HRI community and define some untapped opportunities.

## Conclusion

AI is and will continue to change the landscape of HRI, with MHRI no exception to this. As Responsible AI becomes more integrated within autonomous systems, the level of autonomy for these systems will continue to increase to allow humans to 'command' the systems while expecting reliable execution of intents. MHC1 will be expected in MHRI, defining and shaping system-level requirements from the interface design to the underlying protocols and languages that enable interoperability and information exchange in these systems.

In this chapter, we propose MHC1 as the natural evolution to meaningful human control. MHC1 is not a revolution within the military who have been adopting the principles of mission command, where commanders communicate intent to subordinates who are provided with a greater degree of freedom in their selection of the courses of action required to achieve these intents. In return, commanders expect that the intent is achieved within established bounds, as well as legal and ethical guardrails. MHC1 inherits principles from mission command that apply to robots and where a human commander will communicate their intent information; this time not to a human subordinate but to robots' subordinates. Robots will have the equivalent of a rank, not to anthropomorphise them, but as a symbol of their operational and mission suitability, a symbol for information flow within the command and control (C2) structure, and a symbol for disciplined initiative.

The chapter introduced artificial intelligence, C2 and discussed some of the challenges for MHRI. We then presented a vignette to depict a picture of how mission commands can operate with robotics and autonomous systems. We drew from the vignette lessons and requirements to inform the design of MHC1 and offered a few key considerations for the rationale of using MHC1 to guide the principles needed to integrate MHRI systems.

Before concluding this chapter, it is worth summarising a few future directions in MHRI necessary to cover the needs for MHC1.

In this work and that of (Bunker, 2020), machine applicability principles are broadly explored for specific sets of mission command principles. A more comprehensive treatment is necessary here to understand not only the interrelation of principles with respect to machine applicability, but also identify those principles that may form the basis of an MHC1 implementation across policy, procedural and technical systems of control.

To achieve ‘meaningful’, a few design requirements are necessary. We present below three of the most prominent requirements for mission command in MHRI. These are chosen due to their necessity in enabling mission command and their complexity, which defines the future roadmap for the research.

Requirement 1: Increased autonomy is a necessary condition for MHC1. A robot that can only perform a limited number of actions in very constrained contexts will offer little utility for a commander, despite its offering a greater utility to the troops. It won’t have the capacity to interpret, plan goal and mission, or risk-manage the commander’s intent. To distil this capacity in a robot, AI and autonomy levels need to be uplifted.

Requirement 2: Ethically aligned and responsible design principles need to be applied to ensure that humans can trust the robots. Human-centric design principles will offer ease of usability and the readiness of the robots to operate under the human commander.

Requirement 3: Communication, communication, and communication are necessary for mission command. The robots need to communicate with the commander and vice versa. The human commander needs to understand the performance envelope of the robots before the commander’s intent is issued, but also needs to hear and receive information from the robot to acknowledge understanding the intent and communicate initial plans and counterplans. These communications can occur through multiple modalities. Graphical user interfaces display information visually, voice, gesture, and texting are some key modalities that allow humans and robots to communicate. These modalities need to be designed with mission command in mind by allowing the language used in a commander’s intent to be central in these multi-modal communication channels.

## Acknowledgement

The authors wish to acknowledge and thank Mrs Chloe Wyatt for her gracious illustrations throughout this work. Thank you.

## Disclaimer

All photos in this work have been sourced through the US Department of Defense Visual Information Distribution Service (DVIDS). The following disclaimer is displayed in accordance with the public use notice and limtiations: "The appearance of U.S. Department of Defense (DoW) visual information does not imply or constitute DoW endorsement."